 \documentclass[
journal,
transmag
]{./IEEEtran}

\usepackage{amsmath}
\usepackage{graphicx}% Include figure files
\usepackage{dcolumn}% Align table columns on decimal point
\usepackage{bm}% bold math
\usepackage[]{hyperref}

\usepackage[utf8]{inputenc}
\usepackage[T1]{fontenc}
\usepackage{mathptmx}
\usepackage{xcolor}

\usepackage[per-mode=symbol,separate-uncertainty = false]{siunitx}
\DeclareSIUnit\dBm{dBm}
\DeclareSIUnit\dB{dB}
\DeclareSIUnit\inch{in}
\newcommand{\iu}{{i\mkern1mu}} % imaginary unit

\begin{document}
\title{Spin-wave propagation in metallic CoFe films determined by microfocused frequency-resolved magneto-optic Kerr effect}

\author{\IEEEauthorblockN{Lukas~Liensberger\IEEEauthorrefmark{1,2}\thanks{Corresponding author: L.~Liensberger (Lukas.Liensberger@wmi.badw.de)}, Luis~Flacke\IEEEauthorrefmark{1,2}, David~Rogerson\IEEEauthorrefmark{1,2}, Matthias~Althammer\IEEEauthorrefmark{1,2}, Rudolf~Gross\IEEEauthorrefmark{1,2,3,4}, Mathias~Weiler\IEEEauthorrefmark{1,2}}
\IEEEauthorblockA{\IEEEauthorrefmark{1}Walther-Mei{\ss}ner-Institut, Bayerische Akademie der Wissenschaften, Garching, Germany}
\IEEEauthorblockA{\IEEEauthorrefmark{2}Physik-Department, Technische Universit\"{a}t M\"{u}nchen, Garching, Germany}
\IEEEauthorblockA{\IEEEauthorrefmark{3}Nanosystems Initiative Munich, Munich, Germany}
\IEEEauthorblockA{\IEEEauthorrefmark{4}Munich Center for Quantum Science and Technology (MCQST), Munich, Germany}}


%\markboth{IEEE MAGNETICS LETTERS}%
{}



\maketitle

\begin{abstract}
We investigated the magnetization dynamics of a patterned Co$_{25}$Fe$_{75}$-based heterostructure with a novel optical measurement technique that we call microfocused frequency-resolved magneto optic Kerr effect ($\mu$FR-MOKE). We measured the magnetic field dependence of the dynamical spin-wave susceptibility and recorded a spatial map of the spin-waves excited by a microwave antenna. We compare these results to those obtained on the same sample with the established microfocused Brillouin light scattering technique. With both techniques, we find a spin-wave propagation length of \SI{5.6}{\micro\meter} at \SI{10}{\giga\hertz}. Furthermore, we measured the dispersion of the wavevector and the spin-wave propagation length as a function of the external magnetic field. These results are in good agreement with existing literature and with the employed Kalinkos-Slavin model.
\end{abstract}

%\begin{IEEEkeywords}
%
%\end{IEEEkeywords}

% \fi
\IEEEpeerreviewmaketitle

\section{Introduction}
In the recent years, many advances have been made in utilizing the angular momentum of quantized collective excitations in exchange-coupled magnetic systems (magnons) to transport and store information~\cite{Manipatruni2019}. These magnonic devices are often realized using insulating magnetic materials with low intrinsic damping such as yttrium iron garnet (YIG)~\cite{Talalaevskij2017,HouchenChang2014}. YIG has been found widespread applications in microwave technology or in novel magnonic devices, where especially long spin-wave propagation lengths and high group velocities are desired~\cite{Evelt2016,Hauser2016,Jungfleisch2015,Evelt2018,Collet2017}. The recent discovery of ultra-low magnetic damping in Co$_{25}$Fe$_{75}$~\cite{Schoen2016} will likely furthermore lead to an increased use of metallic magnetic thin films in magnonics. 

The accurate and precise determination of magnetic damping and spin wave propagation lengths is key to designing magnonic devices. To this end ferromagnetic resonance spectroscopy with a vector network analyzer~\cite{Kalarickal2006} is the state of the art technology that is capable of determining the damping characteristics of unpatterned magnetic materials and quantifying the spin-orbit torques in a normal metal/magnetic material heterostructures~\cite{Berger2018}. Although this technique is versatile and powerful due to its phase-resolving capability and frequency accuracy, it is not able to capture the magnetization dynamics locally or in micro-structured devices. In patterned devices, magnetization dynamics are locally probed by optical techniques, such as the microfocused time-resolved magneto-optic Kerr effect ($\mu$TR-MOKE)~\cite{Korner2017,Au2011,Chauleau2014} or microfocused Brillouin Light Scattering ($\mu$BLS)~\cite{Hillebrands1999,Sebastian2015}. 

$\mu$BLS is a particularly widespread technique, which can probe incoherent (thermal) magnetic excitations as well as coherently excited spin-waves, that are typically used for magnonics. However, $\mu$BLS is sensitive to the spin-wave intensity (and not amplitude), making the imaging of spin-wave wavefronts challenging and time-consuming~\cite{Serga2006,Fohr2009}. 

Here, we employ a novel optical technique to study spin dynamics in a spatially resolved manner in the frequency domain. This micro-focused, frequency-resolved magneto-optic Kerr effect ($\mu$FR-MOKE) technique is based on vector network analysis and can directly image spin-wave wavefronts, combining the spatial resolution of $\mu$BLS with the phase-resolving power and frequency accuracy of vector network analyzer-based broadband magnetic resonance spectroscopy.

\section{Setup}

\begin{figure}
\includegraphics{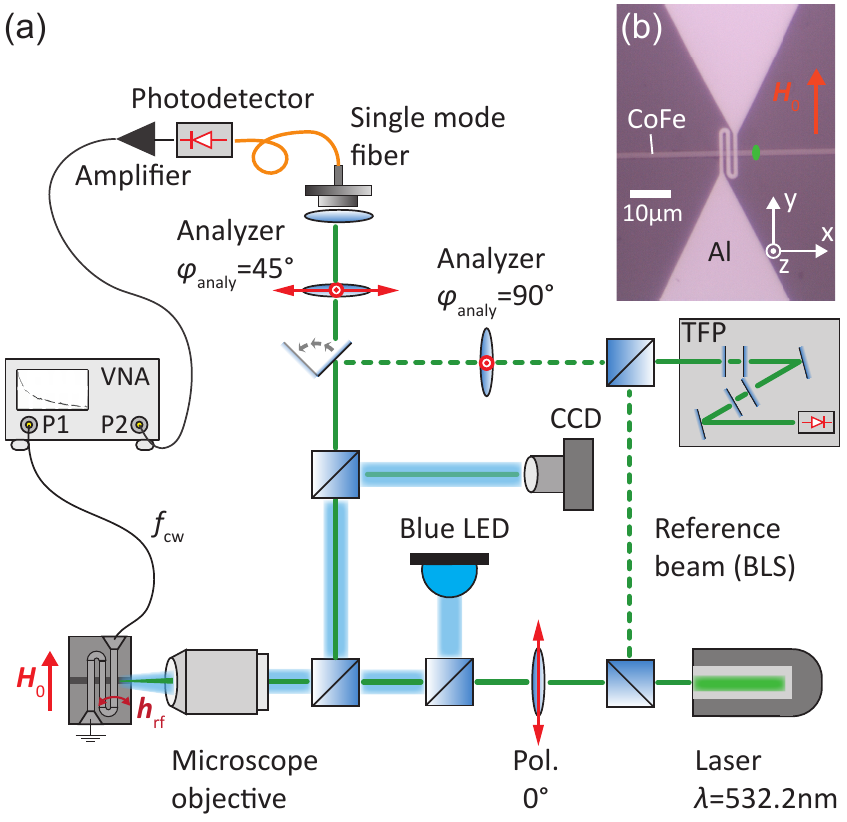}
\caption{(a) Schematic depiction of the $\mu$FR-MOKE and $\mu$BLS setups. The solid green lines indicate the beam path for the FR-MOKE and the dashed lines for the BLS measurement. The inelastically scattered photons are analyzed with a Tandem Fabry-Perot Interferometer (TFP) in the $\mu$BLS measurement. The blue LED and the CCD-camera are used to get an image of the sample and the position of the laser spot on the sample. The red arrows indicate the direction of the polarization axes. (b) Camera image of the investigated sample. The static magnetic field $\bm{H}_0$ is orientated perpendicular to the CoFe-strip (Damon-Eshbach geometry). %The green circle indicates the position of the laser spot for the measurements shown in Fig.~\ref{fig.fig2-fieldsweep}. 
}
\label{fig.fig1-setup}
\end{figure}

The $\mu$FR-MOKE detection principle is based on the frequency-resolved magneto-optic Kerr effect~\cite{Schneider2007,Shaw2009,Nembach2013}. Our setup incorporates micro-focusing and automated image-stabilization to spatially investigate the magnetization dynamics in structured magnonic samples. A schematic depiction of the experimental setup is shown in Fig.~\ref{fig.fig1-setup}(a). A continuous wave laser ($\lambda=\SI{532.2}{\nano\meter}$) is first sent through a Glan-Thompson polarizer and then focused with a microscope objective ($\mathrm{NA}=0.75$) onto the sample, resulting in a diffraction limited optical resolution of about \SI{430}{\nano\meter}. An external static magnetic field $\bm{H}_0$ is applied in the sample plane and a microwave antenna is used for exciting the magnetization dynamics in the thin film sample via GHz  microwave signals. The polarization direction of the back-reflected laser light is modulated at the microwave frequency due to the polar MOKE~\cite{Schneider2007}. The amount of rotation is proportional to the dynamic out-of-plane component $m_z$ of the magnetization. 

To analyze this frequency-modulated polarization rotation, the back-reflected laser beam is sent through a second polarizer (analyzer) which is rotated by \ang{45} with respect to the first polarizer. %The out-of-plane magnetization $m_z$ (and therefore also the laser light polarization) oscillates at the precession frequency. 
The periodic change of the angle of polarization is thus converted into a change of laser intensity. The laser light is then coupled into a single mode fiber and impinges on a fast broadband photodetector with a bandwidth of \SI{25}{\giga\hertz}. After amplification, the photodetector signal is sent to port 2 of a VNA. The VNA also excites the magnetization dynamics at a fixed frequency $f_\mathrm{cw}$ as its port 1 output is coupled into the microstrip antenna and consequently generates an oscillating magnetic field $\bm{h}_\mathrm{rf}$. The VNA phase-sensitively analyzes the signal coming from the photodiode by measuring the complex-valued transmission parameter $S_{21}=V_{2}/V_{1}$ with an IF bandwidth of \SI{1}{\hertz}. Here, $V_1$ is the output voltage applied to the antenna and $V_2$ is the (amplified) output of the photodiode. The applied microwave power is $P=\SI{-5}{\dBm}$ and we confirmed that the magnetic system is in the linear regime for the corresponding excitation field at the antenna. 
Our setup thus probes the complex-valued $m_z$ as a function of frequency, magnetic field, and position.   %although the polar MOKE is compared to longitudinal MOKE larger and therefore the dominating contribution\cite{Hamrle2010}\todo{Maybe we also need to discuss that for the longitudinal MOKE the contributions are averaged out due to the usage of a microscope objective?}. 
In our case, the static magnetic field $H_0$ is applied perpendicular to the magnonic waveguide in the so-called Damon-Eshbach geometry~\cite{Demokritov2001} (see Fig.~\ref{fig.fig1-setup}(b)). 

For the BLS measurements, the back-reflected laser light is sent through a polarizer rotated by \ang{90} with respect to the first polarizer in order to suppress the elastically scattered light. This light is then sent to a (3+3) pass tandem Fabry-Perot interferometer (TFP) where the light is detected by a single-photon detector. A reference beam is used to stabilize the TFP as described elsewhere~\cite{Hillebrands1999}. 

We investigate a patterned thin film based on the low-damping metallic ferromagnet Co$_{25}$Fe$_{75}$~\cite{Schoen2016,Flacke2019}. The magnonic waveguide (width$=\SI{1.8}{\micro\meter}$) and the meander microstrip antenna (width$=\SI{1.2}{\micro\meter}$, gap$=\SI{0.6}{\micro\meter}$) are fabricated using optical lithography and lift-off technique. The aluminum antenna (thickness \SI{50}{\nano\meter}) and the magnonic waveguide are both deposited using DC magnetron sputtering~\cite{Flacke2019,Anders2017} on a Si/SiO$_2$ substrate. For the magnonic waveguide, we deposited Pt(3)/Cu(3)/Co$_{25}$Fe$_{75}$(10)/Cu(3)/Ta(3), where the numbers in the brackets denote the nominal thickness in \si{\nano\meter}. Using in-plane broadband ferromagnetic resonance measurements~\cite{Kalarickal2006} on a reference blanket film, we found a damping of $\alpha_\mathrm{G}=3.94(2)\cdot 10^{-3}$ for this heterostructure. An optical micrograph of the sample is shown in Fig.~\ref{fig.fig1-setup}(b), where the Al-antenna is on top of the waveguide. The antenna is designed to efficiently excite spin-waves with wave vectors up to \SI{6}{\micro\meter^{-1}}.

\section{Experimental results}

\subsection{Comparison of $\mu$FR-MOKE and $\mu$BLS}

\begin{figure}
\includegraphics{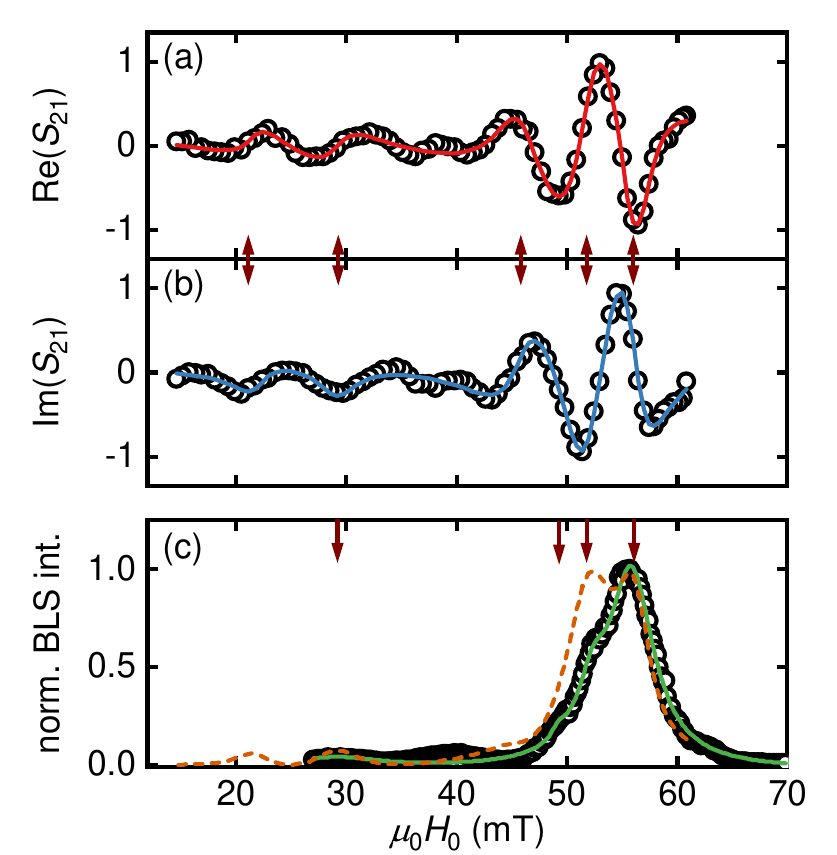}
\caption{Magnetic field dependence of the rescaled (a) real and (b) imaginary part of the $S_{21}$-parameter respectively in the FR-MOKE measurement and (c) the BLS intensity in the Brillouin light scattering experiment. The solid lines are fits according to Eqs.~\eqref{equ.FR-MOKE-S21_freq_fit} and \eqref{equ.BLS-Lorentz} for FR-MOKE and BLS respectively. The red arrows mark the fitted resonance fields and the dashed orange line in (c) corresponds to the fitted $|S_{21}|^2$ from the $\mu$FR-MOKE measurements. }
\label{fig.fig2-fieldsweep}
\end{figure}

First, we measure $S_{21}$ at fixed frequency ($f_\mathrm{cw}=\SI{10}{\giga\hertz}$) and fixed position as denoted by the green spot in Fig.~\ref{fig.fig1-setup}(b) and sweep the static magnetic field $H_0$. The distance of the laser spot from the right side of the antenna is roughly \SI{5}{\micro\meter} so the near-field excitation due to the Oersted field~\cite{Karlqvist1954} of the microwave antenna is strongly suppressed. We compare our results with Brillouin Light Scattering (BLS) measurements~\cite{Hillebrands1999,Sebastian2015} on the same sample. Typical recorded spectra are shown in Fig.~\ref{fig.fig2-fieldsweep}. In order to achieve a comparable signal-to-noise in these measurements, we needed to measure roughly ten times longer in the $\mu$BLS measurements. As the vector network analyzer measures phase-sensitively, the recorded $S_{21}$-parameter is split into its real and imaginary part. As in broadband ferromagnetic resonance, the $S_{21}$-parameter is directly related to the dynamic susceptibility~$\chi$~\cite{Schneider2007,Polder1949} and therefore the $\mu$FR-MOKE data is fitted with
\begin{equation}
S_{21}(H_0) = \sum_i{A_i \chi_i(\omega_\text{cw}, H_0)} + C,
\label{equ.FR-MOKE-S21_freq_fit}
\end{equation}
where $i$ is iterated over the number of resonances ($i\leq 5$), $A_i$ are the complex-valued amplitudes, $\omega_\mathrm{cw}=2\pi f_\text{cw}$ is the angular frequency and $C=C_0 + C_1 \cdot H_0$ is a complex-valued, linear offset to the data. The fits are shown in Fig.~\ref{fig.fig2-fieldsweep}(a) and (b) as solid lines and $C$ has already been subtracted. The resonances correspond to the detection of spin-waves. The resonance with the largest amplitude and a resonance field of $\mu_0H_\mathrm{res}=\SI{56}{\milli\tesla}$ has the smallest wave vector $k$ and spin-waves with larger $k$ are excited for smaller absolute values of $H_0$.

The $\mu$FR-MOKE results are now compared to the established Brillouin Light Scattering technique. Fig.~\ref{fig.fig2-fieldsweep}(c) shows the normalized BLS intensity of the anti-Stokes peak. The detected resonances become apparent as Lorentzian peaks. Therefore, the intensity $I$ is fitted with
\begin{equation}
I(H_0) = \sum_i{\frac{2A_i}{\pi}\,\frac{w_i}{4(H_0-H_{\mathrm{res},i})^2+w_i^2}},
\label{equ.BLS-Lorentz}
\end{equation}
where $H_{\mathrm{res},i}$ are the resonance fields and $w_i$ is the full-width-at-half-maximum linewidth. By comparing the FR-MOKE with the BLS results we find good agreement between these techniques. The small difference in magnetic fields is due to the use of two different magnets for the generation of $H_0$ and technical limitations in field calibration accuracy.

From Fig.~\ref{fig.fig2-fieldsweep}, an important difference between FR-MOKE and BLS becomes evident. The detected MOKE is directly proportional to the dynamic out-of-plane magnetization (polar MOKE, $\propto m_z$) whereas BLS is sensitive to the intensity ($\propto m_z^2$)~\cite{Hamrle2010}. Therefore, the phase information is lost in our BLS measurements. While the phase information can be reconstructed by using an acousto-optic modulator~\cite{Serga2006,Fohr2009}, the BLS signal is fundamentally proportional to $m_z^2$, reducing the sensitivity for the detection of small $m_z$.

\subsection{Spatially resolved spin-wave propagation}

\begin{figure}
\includegraphics{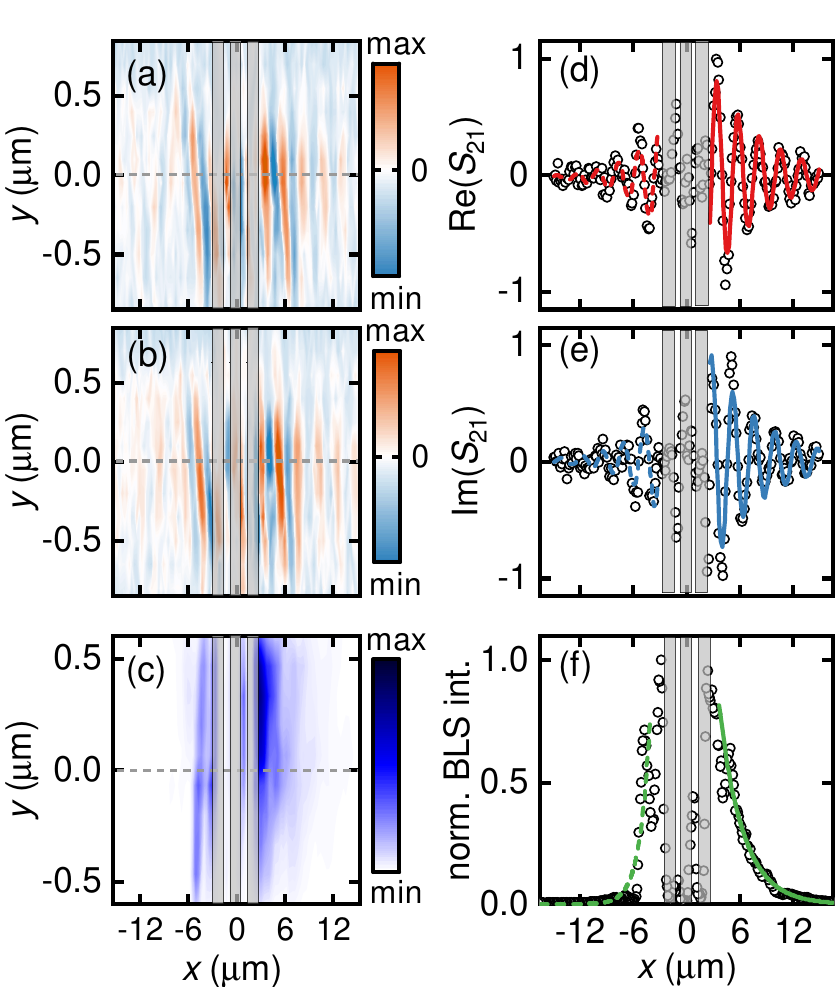}
\caption{Spatial map of the normalized (a) real and (b) imaginary part of the $S_{21}$-parameter (FR-MOKE) and of the (c) normalized BLS intensity measured by scanning the laser over the CoFe-strip at fixed frequency $f_\mathrm{cw}=\SI{10}{\giga\hertz}$ and fixed external magnetic field $\mu_0H_0=\SI{56}{\milli\tesla}$. The gray shaded area indicate the aluminium feedline of the microwave antenna. All measured quantities are shown on a linear scale. (d),(e),(f) Spatial dependence of the $S_{21}$-parameter and the BLS intensity through the middle of the CoFe-strip as indicated by the gray dashed lines in (a), (b) and (c). The solid lines indicate the fits according to Eqs.~\eqref{equ.FR-MOKE-S21_pos_fit} and \eqref{equ.BLS-exp_decay_fit} for FR-MOKE and BLS respectively. }
\label{fig.fig3-spatialscan}
\end{figure}

In Fig.~\ref{fig.fig3-spatialscan}, we present spatially-resolved $\mu$FR-MOKE and $\mu$BLS measurements. Here, the laser spot is scanned in the $xy$-plane with a step width of \SI{150}{\nano\meter} at a fixed microwave frequency of $f_\mathrm{cw}=\SI{10}{\giga\hertz}$ and a fixed static magnetic field of $\mu_0H_0=\SI{56}{\milli\tesla}$ which corresponds to the extracted resonance field of the first spin-wave with the smallest wave vector $k$ in Fig.~\ref{fig.fig2-fieldsweep}. In the FR-MOKE data, the wavefronts of the spin-wave are resolved due to the phase-sensitive measurement. %The total measurement time to record the spatial maps with comparable signal-to-noise was 12~hours for $\mu$FR-MOKE and 12~days for $\mu$BLS. 
With both measurement techniques we observe the non-reciprocity of the spin-wave amplitude due to the antenna non-reciprocity~\cite{Bailleul2003,Nakayama2015}. First, we will focus on the signal to the right side of the antenna, where the signal is larger. 

From the color-coded spatial maps we observe a decay of the amplitude with increasing distance from the antenna. To quantify this decay, we take a line-scan through the middle of the magnonic waveguide as indicated by the dashed gray line in Fig.~\ref{fig.fig3-spatialscan}. These line-scans along the $x$-direction are shown in Figs.\ref{fig.fig3-spatialscan}(d), (e) and (f) for the complex transmission $S_{21}$-parameter and the normalized BLS intensity, respectively. Due to the phase-sensitivity of the FR-MOKE technique we can extract the spin-wave decay length $\xi_\mathrm{sw}$ and the wave vector $k$ from a single measurement. To this end, we fit the real and imaginary part of $S_{21}$ shown in Fig.~\ref{fig.fig3-spatialscan}(d) and (e) simultaneously with
\begin{equation}
S_{21}(x) = A \cdot \exp{\left(-\frac{x}{\xi_\mathrm{sw}}\right)} \cdot e^{\iu k x} + C_0,
\label{equ.FR-MOKE-S21_pos_fit}
\end{equation}
with a complex-valued scaling parameter $A$. %which is an additional fit parameter and accounts for the finite electrical length. 
From the fits, we extract $k=\SI{2.63(1)}{\micro\meter^{-1}}$ and $\xi_\mathrm{sw}=\SI{5.6(4)}{\micro\meter}$. The exponential decay of the BLS intensity shown in Fig.~\ref{fig.fig3-spatialscan}(f) is fitted with~\cite{Demidov2016}
\begin{equation}
I(x) = B \cdot \exp{\left(-2\,\frac{x}{\xi_\mathrm{sw}}\right)},
\label{equ.BLS-exp_decay_fit}
\end{equation}
with a real-valued scaling parameter $B$. The factor 2 in the exponential function is due to the already mentioned sensitivity of the BLS to $m_z^2$. This fit yields $\xi_\mathrm{sw}=\SI{5.4(2)}{\micro\meter}$, in agreement with the extracted FR-MOKE value.

\subsection{Dispersion relation and propagation length}

\begin{figure}
\includegraphics{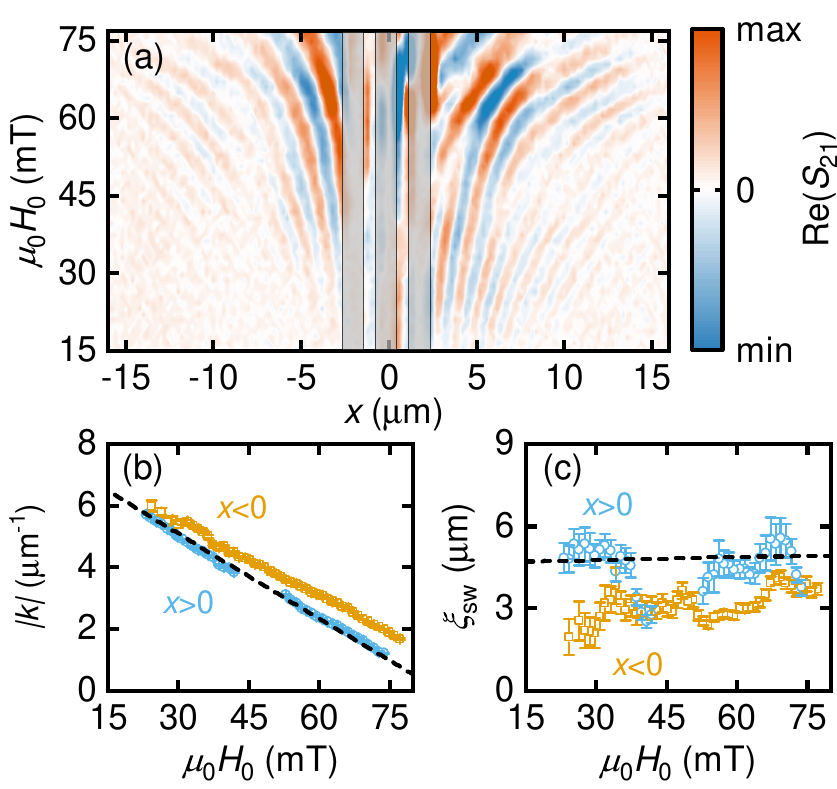}
\caption{Line-scan along the middle of the CoFe strip with a combined magnetic field sweep. (a) Color coded real part of the $S_{21}$-parameter as a function of the spatial coordinate $x$ and the static magnetic field $\mu_0H_0$. (b) Extracted wave vector of the spin-wave $k$ and (c) spin-wave propagation length $\xi_\text{sw}$ as a function of the external magnetic field. The dashed black lines are the results from the Kalinikos-Slavin model (see text).}
\label{fig.fig4-spatialfieldsweep}
\end{figure}

In Fig.~\ref{fig.fig4-spatialfieldsweep} we show line-scans through the middle of the magnonic waveguide at different magnetic fields measured with FR-MOKE at $f_\mathrm{cw}=\SI{10}{\giga\hertz}$. The color-coded spatial-field map shown in Fig.~\ref{fig.fig4-spatialfieldsweep}(a) shows spin-wave wavefronts with changing magnetic field. Again, the spin-wave non-reciprocity can be clearly observed in this measurement scheme. 

We now fit the data in Fig.~\ref{fig.fig4-spatialfieldsweep} for all values of $H_0$ to Eq.~\eqref{equ.FR-MOKE-S21_pos_fit} in order to extract the wave vector $k$ and the spin-wave propagation length $\xi_\mathrm{sw}$ as a function of magnetic field. In Fig.~\ref{fig.fig4-spatialfieldsweep}(b), the wave vector $k$ is shown as a function of the external magnetic field $H_0$ for the positive and negative $x$-axis. %As expected from literature \cite{Korner2017}\todo{Zitat?!} the wave vector $k$ decreases linearly with increasing magnetic field $H_0$. 
In the field range from \SIrange{40}{50}{\milli\tesla} in the positive $x$-direction we could not extract the wave vector. We believe this is due to spin wave interference effects. The largest observable wave vector $k\approx\SI{6}{\micro\meter^{-1}}$ corresponding to a spin-wave wavelength of about \SI{1}{\micro\meter} which is in accordance with the diffraction limited spatial resolution (\SI{430}{\nano\meter}) of our setup. 

The spin-wave propagation length is shown in Fig.~\ref{fig.fig4-spatialfieldsweep}(c). We observe an averaged propagation length of approximately \SI{5.0}{\micro\meter} in the positive $x$-direction and approx. \SI{3.8}{\micro\meter} in the negative direction. Theses results are in good agreement with previous findings~\cite{Korner2017}. The difference in the wave vector $k$ and the spin-wave propagation length $\xi_\mathrm{sw}$ between the positive and negative $x$-direction can be caused by several effects, including antenna non-reciprocity~\cite{Bailleul2003,Nakayama2015}, non-reciprocity of Damon-Eshbach spin-waves and Dzyaloshinskii-Moriya interaction~\cite{Nembach2015}.

To confirm that the extracted wave vectors and spin-wave propagation lengths are in agreement to expectations, we use the Kalinkos-Slavin model~\cite{Kalinkos1989} for the Damon-Eshbach geometry. The spin-wave dispersion reads
\begin{align}
\begin{split}
\omega_\mathrm{r}= \gamma\,& \mu_0\, \sqrt{H_0 + H_\mathrm{d} + H_\mathrm{aniso} + M_\mathrm{s}\, \frac{1-\exp{(-k t)}}{k t}} \\
& \cdot \sqrt{H_0 + H_\mathrm{d} + M_\mathrm{s} \cdot \left(1 - \frac{1-\exp{(-k t)}}{k t}\right)},
\end{split}
\label{equ.Kalinikos_Slavin}
\end{align}
where the saturation magnetization $\mu_0 M_\mathrm{s}=\SI{2.36}{\tesla}$~\cite{Flacke2019}, the demagnetization field $\mu_0 H_\mathrm{d}=-\SI{240}{\milli\tesla}$ and the gyromagnetic ratio $\gamma = g \mu_\mathrm{B}/\hbar$ with the Bohr magneton $\mu_\mathrm{B}$, the reduced Planck constant $\hbar$ and the Land\'{e} factor $g=2.067$ were determined from in-plane broadband ferromagnetic resonance measurements on a reference blanket film. In Eq.~\eqref{equ.Kalinikos_Slavin}, $t=\SI{10}{\nano\meter}$ is the thickness of the CoFe-film. The in-plane anisotropy field $\mu_0 H_\mathrm{aniso}=-\SI{31}{\milli\tesla}$ is used as a free parameter. Solving Eq.~\eqref{equ.Kalinikos_Slavin} numerically for $k$ and using $\omega_\mathrm{r}/2\pi=f_\mathrm{cw}=\SI{10}{\giga\hertz}$, we can extract the wave vector $k$ as a function of the magnetic field $H_0$ as shown in Fig.~\ref{fig.fig4-spatialfieldsweep}(b) as a dashed black line.

The spin-wave propagation length is given by $\xi_\mathrm{sw}=v_\mathrm{g} \tau$ with the group velocity $v_\mathrm{g}=\partial \omega_\mathrm{r}/\partial k$ and the lifetime of the spin-wave $\tau=1/\Delta\omega$. The resonance linewidth is given by $\Delta\omega = \alpha_\mathrm{eff} \mu_0 \gamma\,(M_\mathrm{s}/2+H_0+H_\mathrm{d}+H_\mathrm{aniso}/2)$~\cite{Stancil2009} with the effective damping parameter $\alpha_\mathrm{eff}=\alpha_\mathrm{G} + \gamma \mu_0 \Delta H_\mathrm{inh}/(2 \omega_\mathrm{r})$~\cite{Collet2016}. The inhomogenous linewidth broadening $\mu_0\Delta H_\mathrm{inh}=\SI{1.8}{\milli\tesla}$ is determined from broadband ferromagnetic resonance measurements. We obtain the spin-wave propagation length shown in Fig.~\ref{fig.fig4-spatialfieldsweep}(c) as a dashed black line which is in good agreement with the experimentally extracted values.

\section{Summary}

In summary, we have used a novel optical measurement technique to determine the spin-wave propagation in a structured magnonic waveguide down to the diffraction limit of our setup. Our technique is spatially resolved and phase-sensitive. We demonstrated the capability of the $\mu$FR-MOKE technique by investigating spin-wave dynamics in a patterned Co$_{25}$Fe$_{75}$-based heterostructure. The extracted spin-wave wave vectors and the spin-wave propagation length of \SI{5.6}{\micro\meter} is compatible with earlier findings and with the results obtained from independent $\mu$BLS measurements on the same sample. We modeled the measured wave vector and spin-wave propagation length vs. external magnetic field dependence by using the Kalinkos-Slavin model and find a good agreement with our experimental data.

\section*{Acknowledgment}

We gratefully acknowledge financial support by Deutsche Forschungsgemeinschaft (DFG, German Research Foundation) via projects WE5386/4-1, WE5386/5-1 and AL2110/2-1 and via Germany's Excellence Strategy EXC-2111-390814868. We acknowledge valuable discussions with H.~Huebl. M.W.~thanks T.~Silva, H.~Nembach and J.~Shaw for assistance in designing the experimental setup.

\ifCLASSOPTIONcaptionsoff
  \newpage
\fi



\bibliographystyle{IEEEtran}
%\begin{thebibliography}{1}


%\end{thebibliography}

\nocite{*}
%\bibliography{FRMOKE-CoFe_bib}% Produces the bibliography via BibTeX.



%\end{IEEEbiography}


%\newpage



%\vfill


\end{document}